\documentclass{emulateapj}

\newcommand{\Lya}{Ly$\alpha$}

\newcommand{\kms}{\,km~s$^{-1}$}      % note leading thinspace

\def\h50{\, h_{50}^{-1}}

\newcommand{\lseq}{\mbox{\raisebox{-0.7ex}{$\;\stackrel{<}{\sim}\;$}}}

\def\spose#1{\hbox to 0pt{#1\hss}}
\def\simlt{\mathrel{\spose{\lower 3pt\hbox{$\mathchar"218$}}
     \raise 2.0pt\hbox{$\mathchar"13C$}}}
\def\simgt{\mathrel{\spose{\lower 3pt\hbox{$\mathchar"218$}}
     \raise 2.0pt\hbox{$\mathchar"13E$}}}

\newcommand{\oiii}{[\textrm{O}\textsc{iii}]}
\newcommand{\oii}{[\textrm{O}\textsc{ii}]}
\newcommand{\oiidoub}{[\textrm{O}\textsc{ii}]\ensuremath{\lambda3727,3729}}
\newcommand{\oiilam}{[\textrm{O}\textsc{ii}]\ensuremath{\lambda3727}}
\newcommand{\oiiiv}{[\textrm{O}\textsc{iii}]\ensuremath{\lambda5007}}

\newcommand{\oiiii}{\textrm{O}\textsc{iii}]\ensuremath{\lambda1666}}
\newcommand{\oiiidoub}{[\textrm{O}~\textsc{iii}]\ensuremath{\lambda\lambda4959,5007}}
 
\newcommand{\ha}{\ifmmode {\rm H}\alpha \else H$\alpha$\fi}
\newcommand{\hb}{\ifmmode {\rm H}\beta \else H$\beta$\fi}
\newcommand{\lya}{\ifmmode {\rm Ly}\alpha \else Ly$\alpha$\fi}
\newcommand{\pg}{\ifmmode {\rm P}\gamma \else Pa$\gamma$\fi}
\newcommand{\lyb}{\ifmmode {\rm Ly}\beta \else Ly$\beta$\fi}
\newcommand{\lyg}{\ifmmode {\rm Ly}\gamma \else Ly$\gamma$\fi}
\newcommand{\ciii}{\textrm{C}\textsc{iii}]\ensuremath{\lambda1908}}
\newcommand{\ciiidoub}{\textrm{C}\textsc{iii}]\ensuremath{\lambda\lambda1907,1909}}

\newcommand{\civ}{\textrm{C}\textsc{iv}\ensuremath{\lambda1548,1550}}
\newcommand{\civmed}{\textrm{C}\textsc{iv}\ensuremath{\lambda 1550}}
\newcommand{\heii}{\textrm{He}\textsc{ii}\ensuremath{\lambda1640}}
\newcommand{\oiiiuv}{\textrm{O}\textsc{iii}]\ensuremath{\lambda1661,1666}}

\newcommand{\myemail}{eros.vanzella@oabo.inaf.it}

\shorttitle{Spectroscopy of faint galaxies}
\shortauthors{Vanzella et al.}

\begin{document}

\title{High-resolution spectroscopy of a young, low-metallicity optically-thin  $L=0.02L^{*}$ star-forming galaxy at $z=3.12$ \altaffilmark{$\dagger$}}

\author{\sc E. Vanzella\altaffilmark{1,*},
S. De Barros\altaffilmark{1},
G. Cupani\altaffilmark{2},
W. Karman\altaffilmark{3},
M. Gronke\altaffilmark{4},
I. Balestra\altaffilmark{5,2},
D. Coe\altaffilmark{6},
M. Mignoli\altaffilmark{1},
M. Brusa\altaffilmark{7},
F. Calura\altaffilmark{1},
G.-B. Caminha\altaffilmark{8},
K. Caputi\altaffilmark{3},
M. Castellano\altaffilmark{9},
L. Christensen\altaffilmark{10},
A. Comastri\altaffilmark{1},
S. Cristiani\altaffilmark{2},
M. Dijkstra\altaffilmark{4},
A. Fontana\altaffilmark{9},
E. Giallongo\altaffilmark{9},
M. Giavalisco\altaffilmark{11},
R. Gilli\altaffilmark{1},
A. Grazian\altaffilmark{9},
C. Grillo\altaffilmark{10},
A. Koekemoer\altaffilmark{6},
M. Meneghetti\altaffilmark{1},
M. Nonino\altaffilmark{2},
L. Pentericci\altaffilmark{9},
P. Rosati\altaffilmark{8},
D. Schaerer\altaffilmark{12},
A. Verhamme\altaffilmark{12},
C. Vignali\altaffilmark{7} and \\
G. Zamorani\altaffilmark{1}
 }

\altaffiltext{1}{INAF--Osservatorio Astronomico di Bologna, via Ranzani 1, 40127 Bologna, Italy}
\altaffiltext{2}{INAF - Osservatorio Astronomico di Trieste, via G. B. Tiepolo 11, I-34131, Trieste, Italy}
\altaffiltext{3}{Kapteyn Astronomical Institute, University of Groningen, Postbus 800, 9700 AV, Groningen, The Netherlands}
\altaffiltext{4}{Institute of Theoretical Astrophysics, University of Oslo, Postboks 1029 Blindern, NO-0315 Oslo, Norway}
\altaffiltext{9}{INAF--Osservatorio Astronomico di Roma, via Frascati 33, 00040 Monteporzio, Italy}
\altaffiltext{5}{University Observatory Munich, Scheinerstrasse 1, D-819 M\"unchen, Germany}
\altaffiltext{7}{Dipartimento di Fisica e Astronomia, Università degli Studi di Bologna, Viale Berti Pichat 6/2, 40127 Bologna, Italy}
\altaffiltext{6}{STScI, 3700 San Martin Dr., Baltimore, MD 21218, USA}
\altaffiltext{11}{Astronomy Department, University of Massachusetts, Amherst, MA 01003, USA}
\altaffiltext{10}{Dark Cosmology Centre, Niels Bohr Institute, University of Copenhagen, Juliane Maries Vej 30, DK-2100 Copenhagen, Denmark}
\altaffiltext{8}{Dipartimento di Fisica e Scienze della Terra, Universit\`a di Ferrara, via Saragat 1, 44122 Ferrara, Italy}
\altaffiltext{12}{Observatoire de Gen\`eve, Université de Gen\`eve, 51 Ch. des Maillettes, 1290, Versoix, Switzerland}

\altaffiltext{*}{\myemail}
\altaffiltext{$\dagger$}{Based on observations collected at the European Southern Observatory for Astronomical research in the
Southern Hemisphere under ESO programmes P095.A-0840, P095.A-0653, P186.A-0798.}

\begin{abstract}
We present VLT/X-Shooter and MUSE spectroscopy of an faint
F814W=$28.60\pm0.33$ ($M_{UV}=-17.0$), low mass (\lseq$10^{7} M_{\odot}$) 
and compact ($R_{eff}=62$pc) freshly star-forming galaxy at $z=3.1169$ magnified 
($16\times$) by the {\it Hubble Frontier Fields} galaxy cluster Abell S1063.
Gravitational lensing allows for a significant jump toward low-luminosity regimes, 
in moderately high resolution spectroscopy ($R=\lambda/d \lambda \sim 3000-7400$).
We measured \civ, \heii, \oiiiuv, \ciiidoub, \hb, \oiiidoub,
emission lines with $FWHM\lseq 50$\kms\ and (de-lensed) fluxes spanning the 
interval $1.0\times10^{-19} - 2\times10^{-18} erg s^{-1} cm^{-2}$ at S/N=4-30. 
The double peaked \lya\ emission with $\Delta v(red-blue) = 280$($\pm7$)\kms\ and de-lensed 
fluxes $2.4_{(blue)}|8.5_{(red)}\times10^{-18} erg s^{-1} cm^{-2}$ (S/N=$38_{(blue)}|110_{(red)}$)
indicate a low column density of neutral hydrogen gas 
consistent with a highly ionized interstellar medium as also inferred from the
large \oiiiv / \oiilam\ $>10$ ratio.
We detect \civ\ resonant doublet in emission, each component with $FWHM \lseq 45$\kms, 
and redshifted by $+51$($\pm 10$)\kms\ relative to the systemic redshift.
We interpret this as nebular emission tracing an expanding optically-thin interstellar medium. 
Both \civ\ and \heii\ suggest the presence of hot and massive stars (with a possible faint AGN).
The ultraviolet slope is remarkably blue, $\beta =-2.95 \pm 0.20$
($F_{\lambda}=\lambda^{\beta}$), consistent with a dust-free and
young \lseq20 Myr galaxy.  Line ratios suggest an oxygen abundance 
12+log(O/H)$<7.8$. We are witnessing an early episode of star-formation
in which a relatively low $N_{HI}$ and negligible dust attenuation
might favor a leakage of ionizing radiation.
This galaxy currently represents a unique low-luminosity reference object for future
studies of the reionization epoch with JWST. 
\end{abstract}

%galaxies: evolution}

\section{Introduction}
The epoch of reionization marks a major phase transition of the Universe, during
which the intergalactic space became transparent to UV photons. Determining when
this occurred, the physical processes involved and the sources of ionizing radiation
represents one of the major goals in observational cosmology. The production of 
ionizing radiation is most probably driven by star formation and/or nuclear activity, 
but their relative contribution to the ionizing background is still matter of debate 
\citep[e.g.,][]{fontanot14}. Irrespective of the nature of ionizing radiation, the general 
consensus is that the faint sources are the main producers of the ionizing background
at high redshift (\citealt{wise14,kimm14,madau15}, but see \citealt{sharma16}). 
This implicitly assumes that a not-negligible fraction of
ionizing photons is not trapped in faint sources and escapes. It is therefore important to push
observations toward low luminosity regimes ($L< 0.1L^{\star}$), to investigate
the nature of ionizing radiation and the opacity at the Lyman continuum (LyC, $<912$\AA).
Furthermore, high-ionization
and narrow atomic transitions (like \civ, \heii, \ciiidoub) recently identified
at $z\sim2-3$ and $z>6$ raised intriguing questions about the presence of hot and
massive stars and/or faint nuclear activity \citep[e.g.,][]{stark14,stark15a}
and/or possibly extreme stellar populations \citep{sobral15}. 
Also the large \oiiiv\ rest-frame equivalent width ($> 500-1000$\AA) and
\oiiiv / \oiilam\ ratio ($>5$) recently observed
in relatively bright ($L\simeq L^{\star}$) LyC emitting galaxies is opening promising prospects
for the characterization of reionizing sources at $z>6$ \citep{debarros16,vanzella16,nakajima14,jaskot13,izotov16}. 
A subsequent step is to extend this study to fainter luminosity regimes. 
Here we push observations to
unprecedented luminosities limits ($L \simeq 0.02L^{\star}$, or F814W$\simeq28.60$)
and ask the questions: what is the nature of the ionizing radiation at very faint luminosity/mass domain? 
Are faint sources optically thin at the LyC as expected if they dominate reionization?
Here a detailed study of a faint galaxy is presented, taking advantage of multi-wavelength
photometry available from CLASH \citep{postman12} and Hubble Frontier Fields (HFF)
projects \citep{lotz14,kokom14}\footnote{http://www.stsci.edu/hst/campaigns/frontier-fields/}
 and from low and medium resolution spectroscopy we obtained  at VLT
(VIMOS, MUSE and X-Shooter).

%%%%%%%%%%%%%%%
\begin{figure*}
 \epsscale{1.0}
 \plotone{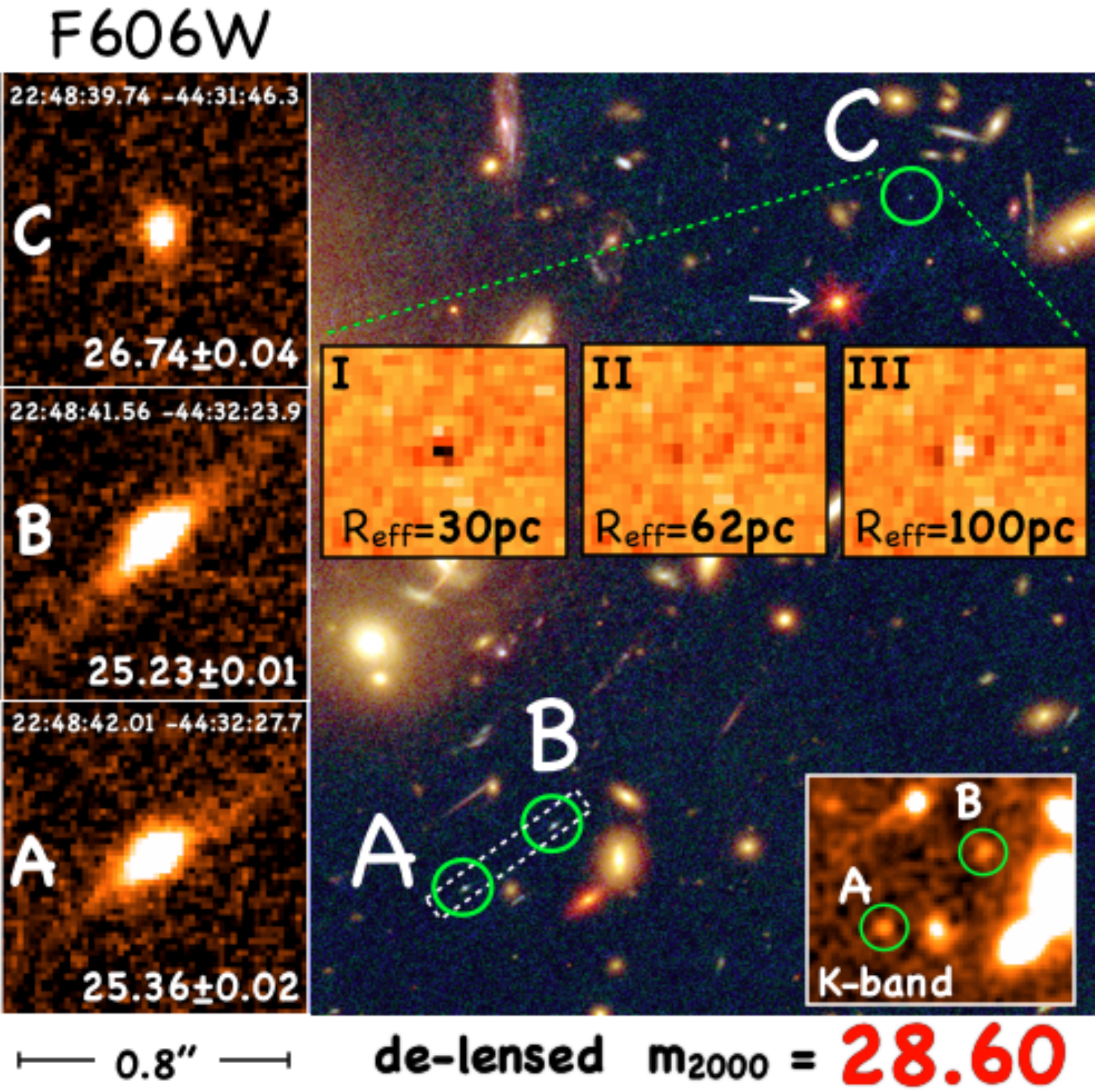} 
\caption{Left: the multiple images A, B and C in the F606W band ($S/N\simeq20-50$), 
the observed magnitudes and coordinates are shown.
Right: the three multiple images are indicated with green circles over the color image ($100\arcsec \times80$\arcsec), 
as well as the orientation and length of the X-Shooter slit (dotted line). 
The insets (\textsc{I,II,III}) show the residuals of the three Galfit models(0.7\arcsec$ \times $0.9\arcsec) 
with different de-lensed $R_{eff}$ calculated on image C (F606W). The arrow indicates the star used as PSF model.
The bottom-right panel shows the images A and B in the VLT/K-band, boosted by nebular \oiiidoub\ lines.
\label{magni}}
\end{figure*}
%%%%%%

\section{Target selection, magnification and X-Shooter observations}

The source ID11 has been selected among a sample of \Lya~emitters
identified with four hours integration with MUSE behind the HFF galaxy cluster 
AS1063 (or RXJ2248, see \citet{karman15}, K15 hereafter) and previously detected
with VLT/VIMOS low resolution spectroscopy ($R=180$) by \citet{balestra13}.

ID11 is a $z=3.1169$ compact galaxy lensed into three images, A, B and C as 
shown in Figure~\ref{magni}.
The A and B images have very similar F814W magnitudes, $25.65\pm0.02$ and $25.57\pm0.02$, respectively,
while the third one (C) is the faintest with F814W$ = 27.06\pm0.04$ magnitude.
Such a geometric  configuration is well reproduced  by the lensing modeling.
The galaxy is close to a caustic on the source plane
and the corresponding critical line lies approximately between the two images A and B on the lens plane.
Among the three images, the faintest one (C) has the least uncertain 
magnification factor, that is estimated to be $\mu_{C}=4.1\pm0.2$ \citep{caminha16a}.
The magnification of the counter-images A and B have been calculated from the observed
flux ratios between C and images A,B, since the three images originate from the same
source (Figure~\ref{magni}).
The resulting magnifications are $\mu_{A}=15.0$ and $\mu_{B}=16.2$ with
errors smaller than 10\%, inferred from the photometry and the more accurate 
estimate of $\mu_{C}$, and are consistent with those derived from lens modeling
by \citet{caminha16a}.
The de-lensed magnitude of the source in the F814W band (probing the continuum
at $\simeq 2000$\AA~rest-frame) is $28.60\pm 0.33$. 

The VLT/X-Shooter \citep{vernet11} observations
of source ID11 have been performed by inserting components A and B in the slit 
(Figure~\ref{magni}). 
Out of 4 hours requested only 2 have been executed.
However, combining the two counter-images
A and B, the equivalent of four hours of integration have been achieved with a
spectral resolution $R$ of 5000, 7350 and 5100 in the three UVB ($\simeq$3000-5600\AA), 
VIS ($\simeq$5500-10000\AA) and NIR ($\simeq$10000-23700\AA) arms, respectively.

Particular care has been devoted to the data reduction of such a faint object.
The data were first reduced using the latest release of the ESO X-Shooter pipeline 
 \citep{modigliani10}. The ESO pipeline 
 produces rectified sky-subtracted spectra of the echelle orders that are useful to 
 determine the position of the two A and B images along the slit. 
 With this information a model of the sky emission on the science exposure has been calculated
 with the technique described in \citet{kelson13}.\footnote{It has been performed with a specific IDL pipeline developed by George Becker.} 
 Wavelength calibration was performed using arc lamp lines (for the UVB arm)
 or the sky emission lines (for the VIS and NIR arms); the resulting r.m.s. was typically 1/10 pixel.
 As a further check, the wavelength positions of the emission lines are 
 fully consistent with what derived from MUSE.
 The combined 1D spectrum was optimally extracted from the 
 wavelength- and flux-calibrated 2D spectra.  
 A resolution-optimized  velocity binning was adopted for the three arms (20, 11, and 19 km s$^{-1}$ for 
 the UVB, VIS, and NIR, respectively).

\begin{figure*}
 \epsscale{1.0}
 \plotone{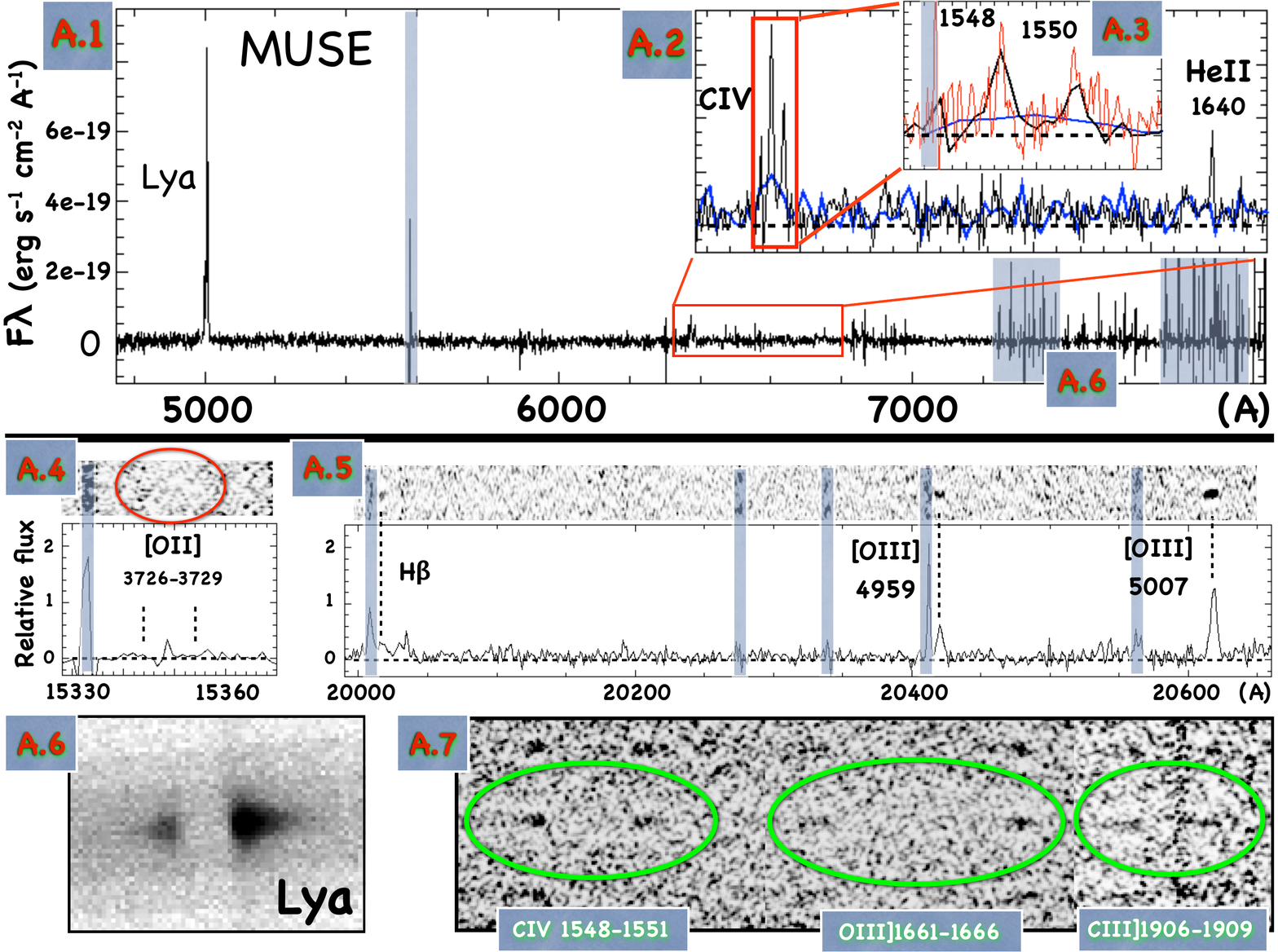}
\caption{{\bf Top part}: the MUSE spectrum is shown (A.1) with indicated the sky emission regions (gray stripes). The zoomed MUSE 
region of \civ\ and \heii\ lines is shown (A.2). A low spectral resolution spectrum is shown ($R=200$, blue line, A.2).
A detail of the X-Shooter \civ\ doublet is shown in A.3,  red line, superimposed to MUSE and VIMOS spectra. 
 {\bf Bottom part}: Two dimensional X-Shooter spectra zoomed around \oiidoub\ (A.4),  \hb\ + \oiiidoub\ (A.5),  \lya\ (A.6), 
  \civ, \oiiiuv\  and \ciiidoub\ (A.7) are shown.
The \lya~line is the sum of the two images A and B computed directly on the raw science frames.
The green ellipses indicate lines after summing the two sky-subtracted images A and B (4hr) and shifting the spectra properly along
the spatial direction.  The emissions above and below the ellipses are the single exposures (2hr).
\label{spectra}}
\end{figure*}

The reduced spectrum and the zoomed \lya\ line are shown in Figure~\ref{spectra} and ~\ref{lya}, respectively. 

%%%%%%%%%%%%%%%%%
%%%%%%%%%%%%%%%%

\section{Results from spectroscopy and multi-band photometry}

\subsection{Spectral properties}

\begin{figure}
 \epsscale{1.0}
 \plotone{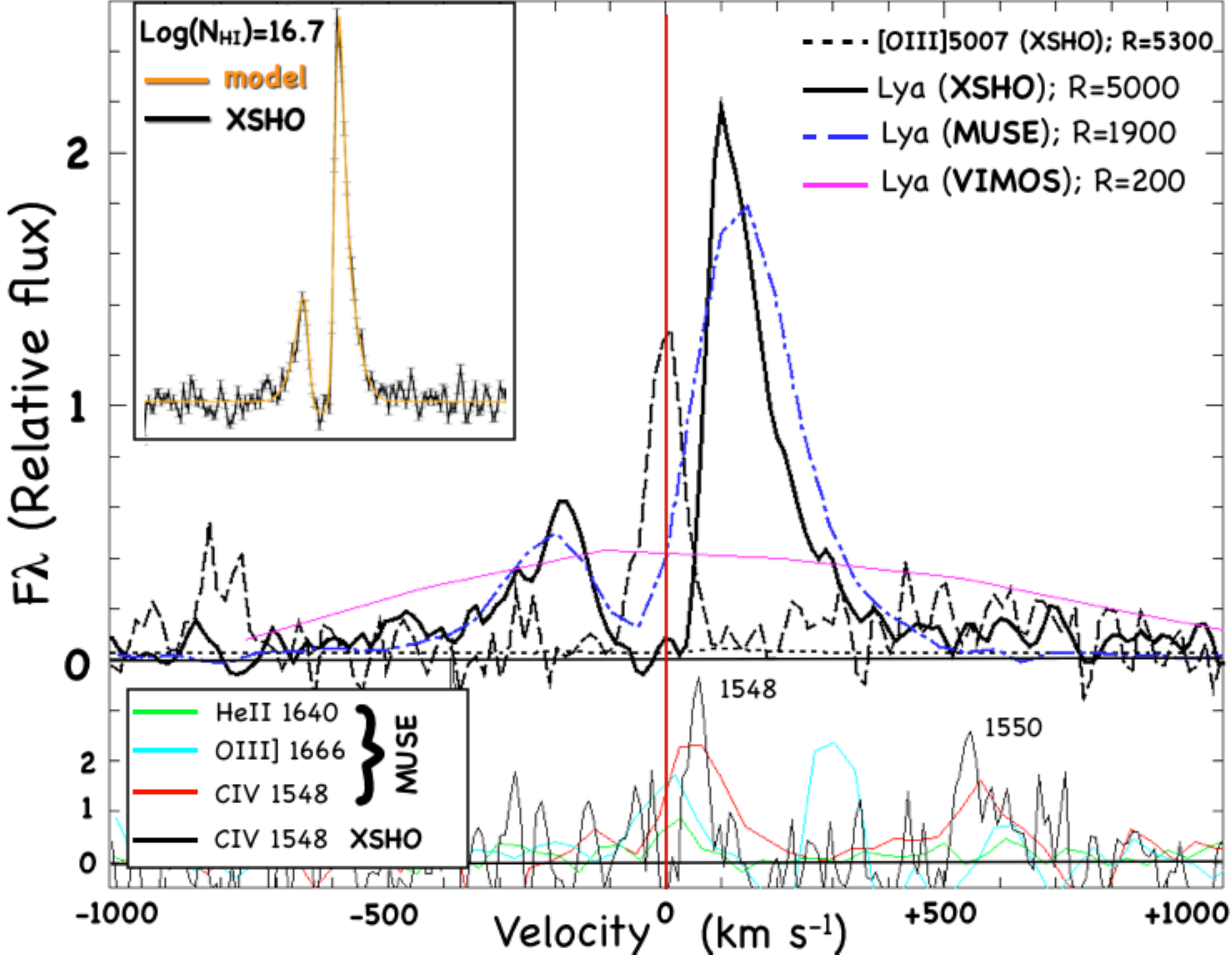}
\caption{Comparison of the most relevant spectral features in the velocity space. 
{\bf Top panel}: the \Lya~line profile is shown for different instruments (VIMOS, MUSE, X-Shooter). 
The \oiiiv\ emission line identified with X-Shooter is shown with a dashed line and marks the assumed
systemic velocity (it has been multiplied by a factor of 4 for graphic purposes).
Both components of the \lya\ line are clearly resolved at the X-Shooter spectral resolution.  
The inset shows an example of \lya\ modeling (orange line). {\bf Bottom panel}: the high ionization emission
lines (as indicated in the legend) are also shown with respect to the systemic velocity. Among them, the only feature showing 
a significant velocity offset  is the \civ\ doublet, with $dv=+50$\kms. \label{lya}}
\end{figure}

The double peaked \Lya~line was initially detected with MUSE (K15). 
The two components of the doublet are resolved in the X-Shooter
spectrum, in which the asymmetric shape with a
trough toward the systemic velocity is evident (see Figure~\ref{lya}).  

As already discussed in K15, faint high ionization emission lines have been
detected (\civ, \heii, \oiiiuv) but not resolved at the MUSE spectral resolution,
placing a limit of $FWHM<100$\kms\ (instrumental corrected). 
As the \lya\ line is the brightest feature in the MUSE data, we used it to 
create a mask to extract the spectra. Although this leads to a higher 
S/N and better flux measurement of \lya, the high-ionization lines do not have 
enough S/N in the outer parts of this mask, and are therefore detected with
lower S/N. We therefore used the mask to extract the \lya\ flux, but 
used a circular 1\arcsec~radius aperture to measure the high-ionization lines, 
and verified that these fluxes are in agreement with the fluxes in the 
full \lya-mask.

While MUSE reaches deeper flux limits with a resolution element $>100$\kms, the X-Shooter
spectral resolution allows us to investigate the width of the features down
to few tens \kms\ and better resolve emission lines close to the sky emission.
We anchor the flux measurements in the VIS arm to the MUSE
ones (e.g. \civ\ doublet), therefore accounting for slit losses and derive X-Shooter
fluxes for other lines missed in MUSE due to sky contamination.
In particular, all the high-ionization lines identified in the X-Shooter spectrum
show a FWHM of \lseq50\kms\ (see Table~1). It is worth noting that an accurate
identification of such narrow lines in high redshift star-forming
galaxies is often compromised in low resolution spectra 
(e.g., panel A.2, of Figure~\ref{spectra}, blue line). 
The equivalent widths of the lines have been estimated
from the line fluxes and the underlying continuum derived from SED fits (see below).

Furthermore, the near-infrared coverage of X-Shooter (up to $2.3\mu m $) allowed us to
detect optical rest-frame emission lines like \hb\ and \oiiidoub\  from which an accurate estimate of the
systemic redshift is performed.
In particular, from \hb, \oiiidoub\ and ultraviolet \oiiiuv\ and \ciii\ lines we derived $z_{syst}=3.1169\pm0.0002$. 
The high ionization emission lines redshifts are consistent with $z_{syst}$, except the \civ\ components
that show a clear velocity shift of $+51$($\pm 10$)\kms\ (Figure~\ref{lya}).
Moreover, the observed \oiii5007/\oii3727 ratio (O32 index) is large ($>10$). 
We discuss below the possible interpretation of such features.

It is worth noting that 
at the given spectral resolution $R=5000-7000$ and S/N of line fluxes, 
the X-Shooter spectrum presented here resembles what a 40-meter class telescope
can achieve in few hours integration time for an unlensed object of the same absolute
magnitude and redshift \citep{evans15,karen14}. 

%%%
%%%%%%

%%%%%%%%%%%%%%
\begin{figure}
 \epsscale{1.0}
 \plotone{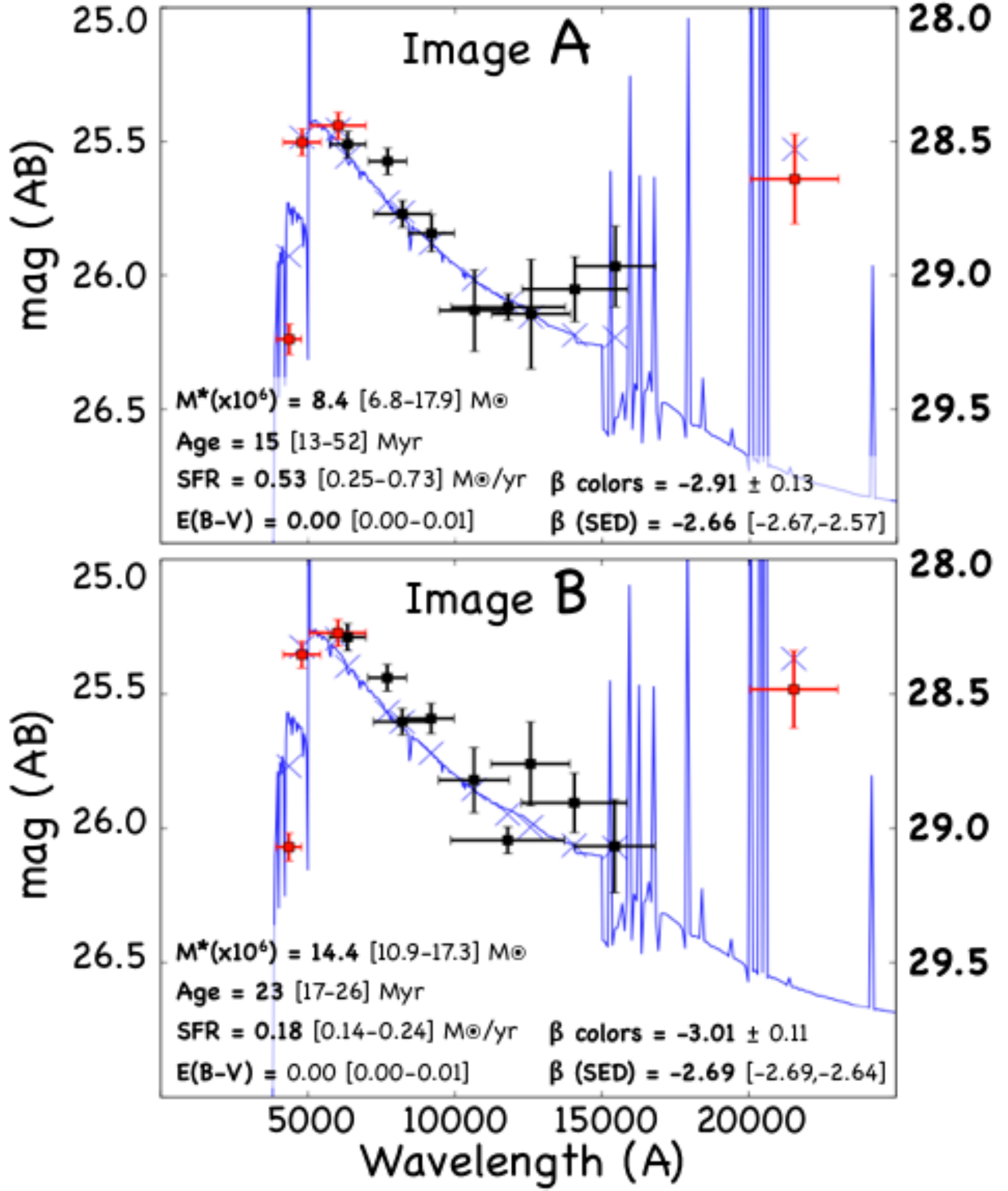} 
\caption{SED fits with Bruzual \& Charlot templates are shown.  The fits have been performed by
both including and excluding the bands significantly contaminated by the IGM and emission lines (red points). The resulting physical
quantities are reported with their 68\% uncertainty. 
Observed and de-lensed magnitudes are reported in the Y-axis on the left and right side, respectively.
\label{SED}}
\end{figure}
%%%%%%

%%%%%%%%%%%%%%%%%%%%%%%%
\begin{deluxetable}{lll}
\tabletypesize{\scriptsize}
\tablecaption{Observed spectral lines. \label{physical}}
\tablewidth{0pt}
\tablehead{
\colhead{Line/vacuum$\lambda$(\AA)} & \colhead{Flux(S/N)[FWHM][EW]} & \colhead{Redshift}}
\startdata
\Lya(blue)${\lambda1215.69}$                                                 & 3.15(38)[104][25]  & 3.1145   \\          
\Lya(red)${\lambda1215.69}$                                                   & 14.53(110)[104][116]  & 3.1184   \\  
$\textrm{C}\textsc{iv}\ensuremath{\lambda1548.20}$          & 0.52(18)[$<45$][7] & 3.1177 \\ 
$\textrm{C}\textsc{iv}\ensuremath{\lambda1550.78}$          & 0.29(10)[$<45$][4] &  3.1175  \\  
$\textrm{He}\textsc{ii}\ensuremath{\lambda1640.42}$         & 0.21(6)[$<100$][3]   & 3.1169$\dagger$\\ 
$\textrm{O}\textsc{iii}]\ensuremath{\lambda1660.81}$          & 0.20(3)[$<45$][3]  & (3.1167)$\dagger$ \\
$\textrm{O}\textsc{iii}]\ensuremath{\lambda1666.15}$          & 0.31(5)[$<45$][5]    & 3.1169 \\
$[\textrm{C}\textsc{iii}]\ensuremath{\lambda1906.68}$         &  0.28(4)[$<45$][6]    &  3.1169 \\
$\textrm{C}\textsc{iii}]\ensuremath{\lambda1908.73}$          &  0.22(2)[$<45$][5]    &  (3.1170)   \\ 
\oii${\lambda3727.09}$                                                               & $< 0.2$               & ---\\
\oii${\lambda3729.88}$                                                               & $< 0.2$                & ---\\
\hb${\lambda4862.69}$                                                               &  0.31(4)[ --][$\simeq110$]      &  (3.1166)  \\
$[\textrm{O}\textsc{iii}]\ensuremath{\lambda4960.30}$         &  0.90(12)[54][$\simeq340$]  &  3.1168  \\ 
$[\textrm{O}\textsc{iii}]\ensuremath{\lambda5008.24}$         &  2.35(33)[51][$\simeq860$]  &  3.1169 \\ 
\tableline
\enddata
\tablecomments{Observed fluxes are reported in units of $10^{-17}ergs^{-1}cm^{-2}$ 
(de-lensed fluxes can be obtained by multiplying the values by 0.06). The S/N, FWHM (instrumental corrected, \kms) and 
rest-frame equivalent width (\AA) are also indicated. The reported fluxes in the wavelength range 1215.68-1660.81\AA\ are
estimated from MUSE. The FWHM, except HeII1640, are estimated from the higher
resolution X-Shooter spectrum.  $\dagger$  indicates redshifts measured form MUSE spectrum.
Redshifts in parenthesis are uncertain due to low S/N.} 
\end{deluxetable}
%%%%%%%%%%%%%%%%

\subsection{Modeling the \Lya~profile}
The separation of the double peaked \lya\ line $\Delta_{peaks} = 280$($\pm7$)\kms\ is 
smaller than commonly found at this redshift.  \citet{kulas12} reported 
typical separations from $400$ up to $1000$\kms\ for brighter $L^{*}$ galaxies. It is instead
slightly lower than the case  reported by \citet{christensen12} in a lensed galaxy at $z=1.83$, 
in which narrow \civ\ emission was also detected. 

The observed small separation 
suggests that $N_{HI}$ is low. Specifically, we modeled the \Lya~structure with the
expanding shell model presented in \citet{gronke15} and described in
Karman et al. (in prep). We refer the reader to those works for details.
Under the model assumptions, a relatively narrow range of $N_{HI}$ is allowed, 
$N_{HI} \simeq 10^{16-18.5} cm^{-2}$ (an example is shown in the inset of Figure~\ref{lya}). 
This result is fully consistent with the analysis of \citet{verhamme15}.
In particular, it is worth noting that , given the estimated range for  $N_{HI}$ a 
leakage of ionizing radiation is also possible
(i.e., $\tau_{LyC}<1$ if $N_{HI}<10^{17.2} cm^{-2}$).
An outflow  velocity of $\simeq 55$($\pm 10$)\kms\ is also derived from the same modeling,
fully consistent with the velocity offset inferred from the \civ\ line doublet (see below).
It is worth mentioning that 
fast outflows ($>100$\kms) can mimic low $N_{HI}$ when inferred from the 
\lya\ profile \citep{verhamme15,schaerer11}.
However, the low velocity expansion derived from the \civ\ doublets
supports a low $N_{HI}$ for this galaxy ($<10^{18.5}cm^{-2}$).

\subsection{The \civ\ doublet and optical oxygen lines}

Another  evidence supporting a transparent medium is the presence of nebular \civ\ emission.
The \civ\ doublet is a resonant transition and is very rarely observed with such narrow components in emission,
possibly due to the low spectral resolution and limited depth of the current spectroscopic
surveys. 
The \civ\  transition is
a combination of stellar P-Cygni emission and broad absorption  \citep[e.g.,][]{kudritzki02},
possible nebular emission, and interstellar absorption superposed \citep{shapley03}. 
In our case the very thin lines ($\sigma_v \leq 20$\kms) 
suggest that the interstellar medium is transparent, allowing the 
\civ\ nebular emission to emerge. This is consistent with the low $N_{HI}$ inferred
from \lya\ modeling mentioned above. 
Furthermore, the doublet is also redshifted by $\simeq 51\pm10$\kms\ 
($z=3.1176$) compared to the systemic velocity ($z_{syst}=3.1169$). 
The measured velocity shift is consistent with the velocity expansion inferred
from the  \lya\ modeling and can be ascribed to thin nebular
emission from a moving medium. 

Optical rest-frame oxygen emission lines also trace the status of the ISM. 
In particular, a large O32 (\oiii5007/\oii3727$>10$) index
has been recently found in a LyC emitter at $z=3.212$ \citep{debarros16},
for which escaping ionizing radiation has been confirmed with HST observations
\citep{vanzella16}.
The source described in this work shows a large O32 index ($>10$),
plausibly linked to a low $N_{HI}$, similarly to what is inferred from
the \Lya~and \civ\ features discussed above.
Such a large O32 index would suggest a density-bounded ISM, 
highly photo-ionized, in which the  \oiilam\ emission is deficient
 \citep[e.g.,][]{jaskot13,nakajima14}.
 
\subsection{A newborn low-metallicity compact galaxy}

Multi-band imaging from the CLASH survey \citep{postman12}, recent
deep HST/ACS observations part of the HFF program (F435W, F606W and F814W)
and additional archival HST data 
have been collected and combined
to produce the photometric SED shown in Figure~\ref{SED} 
\citep[photometry has been extracted following][]{coe15}. 
We also retrieved
and reduced the VLT/HAWKI Ks-band images from the ESO archive (P095.A-0533, P.I. Brammer)
and added it to the SED (see Karman et al. in prep.).
The physical properties have been derived by performing SED fitting with
\citet{bc03} models both on A and B images, and 
accounting for nebular emission by fixing the emission line ratios to the observed
ones \citep{sch09,sch10}.
The fits have been carried out by including/excluding  the 
bands affected by IGM and strong emission lines (\lya\ and \oiiidoub).
The inferred physical quantities agree well in both cases and for the two counter-images (Figure~\ref{SED}). 
The K-band magnitudes dominated by the \oiiidoub\  lines and  are well recovered even when the same
band is excluded from the fit. 
Remarkably, an extremely blue ultraviolet slope is derived for the two
images, from colors directly \citep[e.g.,][]{castellano12} and from the best-fit SED, 
$\beta=-2.95 \pm 0.12 $ and $\beta=-2.7 \pm 0.1 $, respectively (see Figure~\ref{SED}). 
Such a blue shape is compatible with a dust-free
and newborn galaxy with an emergent stellar component of  \lseq20Myr. 
The stellar mass turns out to be $M_{\star}$\lseq$10^{7}M_{\odot}$.
We derived the galaxy metallicity based on the direct $T_e$ method from the
\oiiii / \oiiiv\  ratio, which gives an electron temperature $T_e= 26500\pm2600$~K 
\citep{VM04}. Following \citet{Izotov06} and given the O32$>10$ we derive an oxygen
abundance 12+log(O/H)$<7.8$.
This places the galaxy in the low-mass and low-metallicity region of the mass-metallicity plane
at $z\simeq3$. 

The ultraviolet emission arises from a spatially resolved region.
From the F606W image of  C (S/N$>20$), which has better constrained magnification ($\mu_{C}=4.1\pm0.2$),
the (de-lensed) 
half light radius is $R_{eff}=62$($\pm15$)pc. The $R_{eff}$ and the uncertainty 
have been derived with Galfit \citep{peng10} following the method described in \citet{vanzella15,vanzella16}. 
Figure~\ref{magni} shows three examples of observed--model residuals for three $R_{eff}$ in the F606W band:
30, 62 and 100 parsecs, corresponding to 0.30 (unresolved), 0.55 and 1.00 pixel (1pix=0.03\arcsec).
A similar solution is obtained from the F814W (C), $R_{eff} \simeq 67$pc.
Such a small size, coupled with the aforementioned properties is reminescent of that
observed in a $z\sim3.212$ LyC emitter \citep{vanzella16}, though 
in the present case the source is more than three magnitudes fainter and 4 times smaller.

\section{Discussion and Conclusions}

\subsection{The nature of the ionizing radiation}

The comparison of line ratios like \civmed/\heii, \civmed/\ciii, \oiiii/\heii, \ciii/\heii\ with
models of \citet{feltre16} places the source among the star-forming
galaxies, though still close to the AGN cloud, similarly to the blue galaxies of \citet{stark14}.
Such models are not conclusive for our object, however they do not consider a possible leakage of
ionizing radiation, that could alter the expected ratios both for the UV and optical
rest-frame lines, as for example happens for the O32 index \citep{nakajima14}.
While the high ionization lines are compatible with an
AGN, other properties suggest that the stellar emission is dominating:
the source is spatially resolved in all the HST/ACS images, the very narrow widths
of the involved emission lines ($FWHM<50$\kms)
and the extremely blue slope are not typically observed in AGN-powered objects.
Also the redshifted \civ\ doublet seems to contrast the ubiquitous blueshift
observed in AGN, though at brighter luminosities \citep[e.g.,][]{richards11}.
Therefore, while all of our data can be interpreted with hot and massive stars 
\citep[$T>50000K$,][]{raiter10,grafner15}, only some of them appear to be
consistent with the presence of a faint AGN.

\subsection{A young and naked galaxy: a candidate low-luminosity LyC emitter}

The observed spectroscopic and photometric
properties in such intrinsically faint (F814W(AB)=28.60) galaxy can be measured only as a
result of the factor $\simeq16$ magnification.
The object is a compact ($R_{eff}=62$pc), young (\lseq20Myr), low mass (\lseq$10^{7} M_{\odot}$) 
and dust-free galaxy, with an ionizing source able to generate a
density-bounded condition in the interstellar medium as inferred from the large O32 index.
Such a transparent medium would therefore enable the young stellar component to dominate the
emission and produce the steep ultraviolet slope. 
The redshifted  \civ\ nebular emission is also in line with an expanding optically-thin medium. 
In addition, the very narrow double peaked \lya\ profile ($\Delta_v=280$\kms), the proximity of the red \lya\ peak
to the systemic redshift ($\simeq 100$\kms) and the low velocity outflow
suggest a low $N_{HI}$ ($10^{16-18.5} cm^{-2}$).
Finally, as discussed by \citet{raiter10}, the case of an escaping ionizing radiation
would generate a depression of the nebular continuum that further favors a steepening of the ultraviolet
slope, enhancing the equivalent width of the faint lines like \heii\ or \ciii,
otherwise washed-out by the continuum.

The thinness of the medium is noteworthy in this object and opens  the possibility 
that it is a LyC emitter. Irrespective of possible LyC leakage, the analysis addresses for the first time a
still unexplored luminosity and mass domain, and provide a unique reference lower redshift analog
to the higher redshift blue sources  ($z>6$) at similar luminosities,
believed to be the main actors during reionization \citep{castellano16,bouwens15,atek15}.
It will be crucial to extend the analysis to a statistically significant sample and
fainter luminosity limits.

Even though these extremely blue galaxies could be rare at $z\sim3$, it might not be the case at $z>6$, as
the $\beta$-luminosity relations of \citet{bouwens14} seems to indicate. In particular,
at $z\sim7$ an average $\beta \simeq -2.8$ is expected at the luminosity probed here
($M_{UV}=-17.0$,  see also \citealt{finkelstein12}).
In particular the \civ\ doublet discovered at $z=7.04$ by \citet{stark15b} would be
interesting to compare with the source discussed here.  

Finally, we remark that the initial phases of star-formation as observed here
offer the opportunity to test models of galaxy formation and photoionization effects
in low-mass objects for the first time.

\acknowledgments
We thanks the anonymous referee for a helpful report. 
We acknowledge G.Becker for the valuable support on the X-Shooter data-reduction.
We thanks R.Amorin for discussions about the metallicity of the object
and M.Bellazzini, E.Carretta for useful discussion.
Part of this work has been funded through the PRIN INAF 2012.
MM acknowledges support from PRIN-INAF2014 1.05.01.94.02,  ASI/INAF/I/023/12/0, and
MAECI (US16GR08).

\end{document}